# Universal presence of time-crystalline phases and period-doubling oscillations in one-dimensional Floquet topological insulators

Yiming Pan[1†], Bing Wang[2]

1. Department of Physics of Complex Systems, Weizmann Institute of Science, Rehovot 76100, ISRAEL
2. National Laboratory of Solid State Microstructures and School of Physics, Nanjing University, Nanjing 210093, CHINA

†Correspondence and requests for materials should be addressed to Y.P. (yiming.pan@weizmann.ac.il).

**Abstract**

In this work, we reported a ubiquitous presence of topological Floquet time crystal (TFTC) in one-dimensional periodically-driven systems. The rigidity and realization of spontaneous discrete time-translation symmetry (DTS) breaking in our TFTC model require necessarily coexistence of anomalous topological invariants (0 modes and $\pi$ modes), instead of the presence of disorders or many-body localization. We found that in a particular frequency range of the underlying drive, the anomalous Floquet phase coexistence between 0 and $\pi$ modes can produce the period-doubling (2T, two cycles of the drive) that breaks the DTS spontaneously, leading to the subharmonic response ($\omega/2$, half the drive frequency). The rigid period-2T oscillation is topologically-protected against perturbations due to both non-trivially opening of 0 and $\pi$-gaps in the quasienergy spectrum, thus, as a result, can be viewed as a specific "Rabi oscillation" between two Floquet eigenstates with certain quasienergy splitting $\pi/T$. Our modeling of the time-crystalline 'ground state' can be easily realized in experimental platforms such as topological photonics and ultracold fields. Also, our work can bring significant interests to explore topological phase transition in Floquet systems and to bridge the gap between Floquet topological insulators and photonics, and period-doubled time crystals.

Spontaneous symmetry breaking plays a profound role in modern physics, leading to a variety of condensed states of matter and fundamental particles. In analogy with conventional crystals that break the spatial translation symmetry, F. Wilczek [1-3] proposed the intriguing notion of "time crystal" that spontaneously breaks the continuous time-translation symmetry. Although the possibility of the spontaneous time-translation symmetry breaking in thermal equilibrium was ruled out by a no-go theorem [4-7], Floquet time crystals that break the discrete time-translation symmetry in periodically driven systems proposed in [8-15] attracted intense attention. Since the external driving would heat a closed system to infinite temperature and eventually breaks the time-crystalline phase, disorder-induced many-body localization (MBL) is necessarily proposed to stabilize the long-range time-crystalline phase [9-19]. More interestingly, the expectation values of observables of classical or quantum systems exhibit later-time oscillations with multiple periods than that of the underlying drive [20-22].

Many recent Floquet (or discrete) time crystals (FTCs or DTCs) have been explored both theoretically and experimentally in interacting spin chain models [23-26], ultracold atoms [27, 28] or dipolar systems [29-32]. For instance, these systems are similarly designed based on the periodically-driven MBL phases, e.g., the spin glasses (SGs) in the transverse field Ising model (TFIM) [9, 12], with fully parameter-tunable disorders and interactions. This MBL model enables us to realize a spin-glass phase (e.g., 0SG [9]) with two degenerated and ordered states, such as $|\uparrow\downarrow\uparrow\uparrow\uparrow\downarrow\uparrow \cdots\rangle$ and $|\downarrow\uparrow\downarrow\downarrow\downarrow\uparrow\downarrow \cdots\rangle$. These two SG states are many-body localized and ordered by violating the Ising parity symmetry defined as $X = \prod_i \sigma_i^x$. We can construct a Floquet evolution operator $U_F \equiv exp\left(-i \int_0^T H(t)\, dt\right) = X\, e^{-iH_{SG}t_1}$, characterizing the stroboscopic dynamics within one period of the drive (T), where $H_{SG}$ is the practical spin-glass MBL Hamiltonian [9] and $t_1 < T$. Considering the Ising symmetry of the ordered spin-glass states, we obtain the ensemble eigenstates of our constructed Floquet evolution operator, i.e., $|\{\pm\sigma\}\rangle = (|\uparrow\downarrow\uparrow\uparrow\uparrow\downarrow\uparrow \cdots\rangle \pm |\downarrow\uparrow\downarrow\downarrow\downarrow\uparrow\downarrow \cdots\rangle)/\sqrt{2}$, because the even and odd eigenstates of $X|\{\pm\sigma\}\rangle = \pm|\{\pm\sigma\}\rangle$ are constructed and the superposition states $|\{\pm\sigma\}\rangle$ are still the degenerated eigenstates of $H_{SG}$. As a result, any initial physical superposition state of the Floquet-MBL eigenstates experiences as like $U_F(NT)(\alpha|\{\sigma\}\rangle + \beta|\{-\sigma\}\rangle) = e^{-i\phi(NT)}(\alpha|\{\sigma\}\rangle + (-1)^N \beta|\{-\sigma\}\rangle)$, where $\phi(NT)$ is the accumulated phase after N cycles of the repeats and the coefficients are normalized ($|\alpha|^2 + |\beta|^2 = 1$). The driven spin-glass system exhibits emergent subharmonic oscillations with twice the periods (2T); thus, this stroboscopic

behavior is dubbed as the "$\pi$SG/DTC"[9, 12, 20, 21, 33]. Moreover, from the perspective of quasi-energy spectrum, the sign difference between the Floquet eigenstates $|\{\pm\sigma\}\rangle$ after evolving one cycle implies that the quasienergy difference $|\varepsilon_+ - \varepsilon_-| = \pi/T$ is one of the necessary conditions to achieve the time-crystalline behaviors in Floquet many-body localized systems.

On the other hand, Floquet time crystals do not have to necessarily require many-body localization as a prerequisite. The realizations of time crystals without MBL have also been demonstrated recently, such as a clear FTC model completely without disorders in a strongly interacting regime [27], and an experiment on nitrogen-vacancy centers, in which the FTC was formed regardless of the delocalization by three-dimensional spin-dipolar interactions [21]. Nevertheless, a robust FTC without many-body interaction or disorders has not been achieved [34, 35].

To carry out the rigid time-crystalline phase in a clean non-interacting Floquet system, we are inspired by involving the interplay between topological protection from topological insulators and photonics [36-38] and periodically-driven protocols from Floquet engineering [39, 40]. With the assistance of Floquet engineering, we can obtain the anomalous chiral edge modes in periodically driven systems that lead to anomalous bulk-edge correspondence [41-49]. These anomalous Floquet phases exhibit localization and robustness as edge states or domain walls in the presence of perturbations (fluctuations, disorders, and even weak interactions). As a result, they provide an excellent mechanism to realize the FTCs.

Here, in this present work, we firstly reported a Floquet time crystal in the non-interacting situation that established a stable period-2T oscillation - the 'fingerprint' to experimentally observe a quantum time-crystalline phase - due to the topological protection from nontrivial gap opening in quasienergy spectrum. Our one-dimensional model of FTC is primarily based on Floquet topological phases [41, 42, 49], named the zero-energy edge modes (0 modes) and $\pi$-energy edge modes ($\pi$ modes), which have been extensively studied in periodically driven systems [34, 41-46]. For instance, in the driven Su-Schrieffer-Heeger model as we studied previously [43], we found that in a specific drive frequency region where the two topological phases coexist with quasienergy difference given by $|\varepsilon_\pi - \varepsilon_0| = \pi/T$, the system exhibits a persisting period- 2T oscillation which breaks the underlying discrete time-translation symmetry. We dubbed this period-doubling phenomenon "topological Floquet time crystals (TFTCs)," since in our Floquet time-crystalline model, the topological protection inherited from topological phase coexistence instead of many-body interaction and disorders is of essence to stabilize the long-range subharmonic response($\omega/2$), and robustness against perturbations [50, 51]. Clearly, our results suggest an exciting field of studying time crystals in non-interacting topological Floquet systems, which can be easily implemented in experiments, such as ultracold atoms and topological photonics.

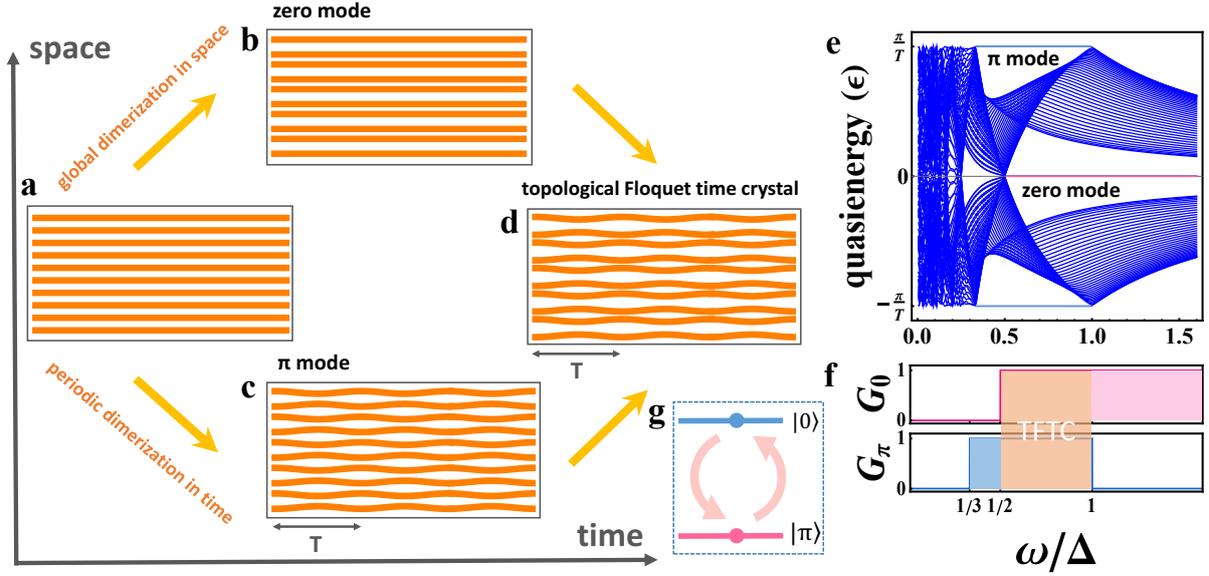

Fig. 1: Floquet engineering of topologically-protected discretized time crystals. (a) a crystal in space and time. (b) the Su-Schrieffer-Heeger (SSH) model is structured when a crystal is spatially dimerized, generating topological 0 modes along the boundaries. (c) the driven SSH model is made when the crystal periodically dimerizes in time and space, expecting that $\pi$ modes would oscillate with time locally on the boundaries. (d) a prototypical driven SSH model is structured by both spatial and temporal dimerization, to produce the anomalous phase coexistence between 0 modes and $\pi$ modes. (e) the quasienergy spectrum corresponding to the driven SSH structure (d) is calculated from the one-cycle time evolution operator $U(T)$ (see Appendix A1). The $\pi$ mode appears at the drive frequency region $(1/3, 1)$, and the 0 mode appears at the region $(1/2, \infty)$. Correspondingly, (f) the topological invariants of 0 and $\pi$ modes is given by the gap invariants $G_0$ and $G_\pi$, respectively. The two protected modes coexist in the region $(1/2, 1)$. (g) depicts the "Rabi" oscillation between 0 and $\pi$ quasienergy levels, due to $|\varepsilon_\pi - \varepsilon_0| = \pi/T$, with the demonstration of subharmonic oscillation with two periods of the drive. The Floquet phase coexistence yields a topological Floquet time crystal (TFTC)

with the spontaneous broken of discrete time-translation symmetry. The relevant parameters in our calculations are $N = 80$, $\kappa_0/\Delta = 0.25$, $\delta\kappa_0/\Delta = 0.06$, $\delta\kappa_1/\Delta = 0.12$, and $\theta = 0$, $\beta_0 = 0$.

**The topological invariants in driven SSH model** - Our modeling of topological Floquet time crystals is, as a simplest example to be demonstrated, based on the prototypical driven Su-Schrieffer-Heeger (SSH) model, which consists of both global dimerization and periodical dimerization in space and time, as schematically depicted in Fig.1a-1d. To realize period-2T FTCs, we have to introduce two topological edge states in our driven SSH model. Fig.1b and 1c depicted the typical structure and drive configurations on how to construct the topological lattices which hold nontrivial edge states, i.e., 0 modes and $\pi$ modes. 0 modes of conventional SSH model (Fig. 1b) have been widely observed in a number of experimental platforms [52-55], while $\pi$ modes of driven SSH model (Fig. 1c) was recently achieved in microwave waveguide arrays [43, 56]. Intuitively, our dimerized SSH model of interest is supposed to coexist both the topological modes by combing both the two kinds of dimerization configurations (Fig. 1d). Since the quasienergy difference between the two modes is given by $|\varepsilon_\pi - \varepsilon_0| = \pi/T$, the topological coexistence can allow the superposition of two Floquet eigenstates that naturally exhibits later-time oscillations with two periods (2T) in system than that of the underlying drive [50, 51]. Mathematically, our period-2T Floquet time crystal based on the driven SSH model (Fig. 1d) is described by the following Hamiltonian,

$$H = \sum_{i=1}^{N} \beta_0 c_i^\dagger c_i + \sum_{i=1}^{N-1} [\kappa_0 + (-1)^i (\delta\kappa_0 + \delta\kappa(t))] c_i^\dagger c_{i+1} + h.c., \qquad (1)$$

where $c_i^\dagger$ and $c_i$ are the creation and annihilation operators at the site $i$ (or the $i^{th}$ waveguide), $N$ is the total number of lattice sites, $\beta_0$ is the constant on-site potential or chemical potential (or propagation constant in a photonic system). The second off-diagonal term in Eq.1 represents the nearest-neighbor hopping, in which $\kappa_0$ is the constant coupling strength, $\delta\kappa_0$ is the globally dimerized staggered coupling strength, and $\delta\kappa(t) = \delta\kappa_1 \cos(\omega t + \theta)$ is the periodically dimerized staggered coupling strength with $\delta\kappa_1$ being the amplitude of the time-periodic coupling, $\omega = 2\pi/T$ the drive frequency and $\theta$ the initial phase of the drive. All four models, as depicted in Fig. 1a-d can be analytically demonstrated with the driven SSH Hamiltonian (Eq.1). For example, as a reference without any dimerization $\delta\kappa_0 = \delta\kappa(t) = 0$, we can obtain its energy dispersion $E_0 = \beta_0 + 2\kappa_0 \cos(k)$ with $a = 1$ being the normalized lattice constant. The local energy bandwidth of this trivial space crystal (Fig. 1a) is thus given by $\Delta = 4\kappa_0$.

The dynamic process of the time-periodic system can be described by the evolution operator $U(t,t_0) = \hat{T}e^{-i\int_{t_0}^{t} H(t')dt'}$ ($\hat{T}$ denotes time ordering), which is the solution of Cauchy problem $i\partial_t U(t,t_0) = H(t)U(t,t_0)$ with the initial value $U(t_0,t_0) = Id$. Due to the periodicity of the evolution operator $U(t_2,t_1) = U(t_2+T,t_1+T)$, we choose the original time $t_0 = 0$ and set $U(t) = U(t,0)$ and then $U(t+nT) = U(t)[U(T)]^n$ with $t \in [0,T]$. The stroboscopic evolution can be fully described by the Floquet operator $U(T)$ from which we can define the effective Hamiltonian $H_{eff} \equiv \frac{i}{T} \ln U(T)$ with $T = 2\pi/\omega$ being the period of the drive. The quasienergy spectrum ($\epsilon$) is thus calculated via the eigenvalue analysis of the effective Hamiltonian $H_{eff}$ and then is presented in Fig. 1e, where the energy-related coupling parameters are normalized with bandwidth $\Delta$. Because of the time periodicity of the drive ($H(t+T) = H(t)$), the quasienergy $\epsilon$ lies in the spectrum ranging from $-\pi/T$ to $\pi/T$, which is denoted as the Floquet-Brillouin zone [39-41]. We can observe that from the quasienergy spectrum, the $\pi$ modes occur in the frequency region $1/3 < \omega/\Delta < 1$, while the 0 modes exist in the region $\omega/\Delta > 1/2$. Thus, we can expect that the topological phase coexistence lies in a certain intersection region, $1/2 < \omega/\Delta < 1$.

These edge states corresponding to the quasienergy spectrum (Fig. 1e) are topologically protected, and we adopt the definitions of gap invariants for quantum periodically-driven systems proposed in [42, 49], and calculate its topological invariants $G_0$ and $G_\pi$ separately for 0 gap and $\pi$ gap in the energy-momentum space (see Appendix A3). Fig. 1f shows that the nontrivial invariants $G_0 = 1$ appears at the frequency region $\omega/\Delta > 1/2$ for 0 modes and $G_\pi = 1$ at the frequency region $1/3 < \omega/\Delta < 1$ for $\pi$ modes, while the rest frequency region is trivial with $G_0 = 0$ or $G_\pi = 0$. Indeed, these two gap invariants have been widely studied in many physical and optical systems [41-45, 49], in which both the presence of two gap invariants indicates the anomalous bulk-edge correspondence in Floquet systems. [41, 43, 49]. As a result, we pointed out that both $\pi$ modes and 0 modes we found in the quasienergy spectrum are topologically protected by the chiral (sublattice) symmetry.

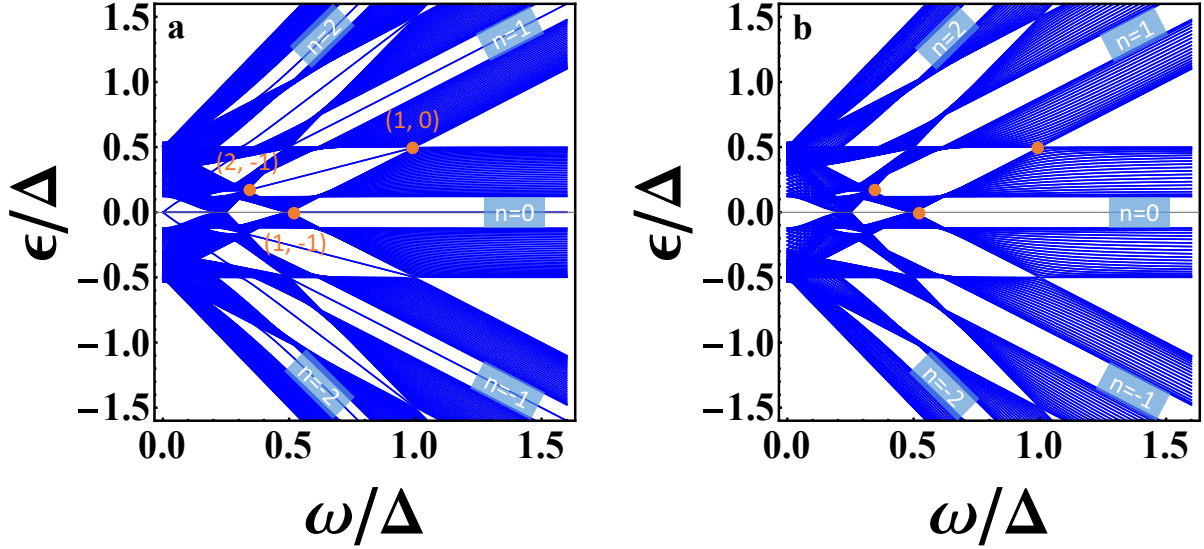

Figure 2: (a)The quasienergy spectrum of Floquet time crystal with the truncated artificial photon bands under open boundary condition(OBC). The driven condition of $\pi$ modes comes from band crossings of two photon bands $(0,1)$ and two bands $(2,-1)$; while the driven condition of 0 modes comes from the crossing of two photon bands $(1,-1)$. These photon band crossings can explain the driven phase transition points $1/3$, $1/2$, and $1$. (b) The quasienergy spectrum of Floquet time crystal with the truncated artificial photon bands under periodic boundary condition(PBC). In this case, the spectrum with imposing PBC shows the same spectral structure and gap opening as that with OBC, expect for no emergent isolated modes laying in the gaps. Here we address that the two gaps nontrivially only open at the frequency region $(1/3, 1/2)$ because of the Floquet photon band crossing between both the bands $n = 2$ and $n = 1$ with the band $n = -1$, respectively.

**Quasi-energy spectrum of Floquet time crystals -** For a time-dependent system with the Hamiltonian $H(t + T) = H(t)$, Floquet theory states that the solution of the Schrödinger equation of the system $(H(t) - i\partial_t)|\psi(t)\rangle = 0 (\hbar = 1)$ has the form $|\psi(t)\rangle = e^{-i\varepsilon t}|u(t)\rangle$, where $H_F = H(t) - i\partial_t$ is the Floquet Hamiltonian, $|\psi(t)\rangle$ is the Floquet state, $\varepsilon$ is the quasienergy, and $|u(t)\rangle$ is the time-periodic Floquet mode. Then the time-dependent Schrödinger equation is mapped to the eigenvalue equation $(H(t) - i\partial_t)|u(t)\rangle = \varepsilon|u(t)\rangle$. Because of the time periodicity, it is convenient to consider the composed Hilbert space $\mathcal{S} = \mathcal{H} \otimes \mathcal{T}$ for the basis $\{|u_{n,\alpha}\rangle e^{in\omega t}\}$. $\mathcal{H}$ is the usual Hilbert space with a complete set of orthonormal basis denoted by $\alpha$, and $\mathcal{T}$ is the space of time-periodic functions spanned by $e^{in\omega t}$ with integer $n$ denoting the $n^{th}$ Floquet replica (artificial photon band). In this space, the quasienergy is given by $\sum_m (H^{(n-m)} + n\omega\delta_{m,n})|u_{m,\alpha}\rangle = \varepsilon_{n,\alpha}|u_{n,\alpha}\rangle$, where $H^{(n)} = \frac{1}{T}\int_0^T e^{-in\omega t} H(t)dt$ is the $n^{th}$-order Fourier component of $H(t)$ with $\omega = 2\pi/T$ and $H(t) = \sum_n e^{in\omega t} H^{(n)}$. The composed scalar product in $\mathcal{S}$ is defined by $\langle\langle\cdots\rangle\rangle = \frac{1}{T}\int_0^T \langle\cdots\rangle dt$, which gets rid of the time dependence.

From the quasienergy spectrum calculated by the Floquet Hamiltonian $H_F$, as shown in Fig. 2a, 2b with imposing open boundary condition(OBC) and periodic boundary condition(PBC) respectively, we can see that under OBC, the photon bands (i.e., the Floquet replicas) $n = 2$ and $n = -1$ cross at the frequency $\omega/\Delta = 1/3$ and the photon bands $n = 1$ and $n = 0$ cross at the frequency $\omega/\Delta = 1$. In the frequency region $1/3 < \omega/\Delta < 1$, $\pi$ mode exists. The 0 mode exists in the frequency region $\omega/\Delta > 1/2$, and the gap starts to open at frequency $\omega/\Delta = 1/2$ where the photon bands $n = 1$ and $n = -1$ cross. There is a particular intersection region $1/2 < \omega/\Delta < 1$, in which the two edge modes coexist. On the contrary, the spectrum with imposing PBC shows the same spectral structure and 0- and $\pi$- gap opening as that with OBC, expect for no emergent isolated modes laying in the gaps. The comparison between the two spectrum indicates the isolated 0 and $\pi$ modes appear locally at the boundary and its underlying topological protection.

Notice that we can calculate the quasienergy spectrum both by the Floquet Hamiltonian $H_F$ in the differential form and by the effective Hamiltonian $H_{eff}$ in the integral form, alternatively, but they are physically equivalent. The Floquet Hamiltonian $H_F$ can be transformed to time-independent effective Hamiltonian $H_{eff}$ by the unitary rotation $V^{-1}(t)H_F(t)V(t) = H_{eff}$, where $V(t) = U(t)e^{iH_{eff}t}$ is the periodized evolution operator that connects to micromotions. (see Appendix A2)

**Period-2T oscillation in topological phase coexistence** – Now, we can explain how the rigid period-2T oscillation surprisingly emerges. Suppose we prepare a superposition of Floquet eigenstates whose phases wind at different rates, for example, $|\phi_\pm(t)\rangle = e^{-i\varepsilon_0 t}|0\rangle \pm e^{-i\varepsilon_\pi t}|\pi\rangle$ where the Floquet eigenstates $|0\rangle$, $|\pi\rangle$ represent the localized 0 modes and $\pi$ modes on the boundaries experiencing different phases under the time evolution operator $U(T)$ over one cycle. The superposition state (i.e., cat state) $|\phi_\pm(t)\rangle$ would repeat with undergoing a 2T period longer than that of the drive, that is, $|\phi_\pm(t + 2T)\rangle = |\phi_\pm(t)\rangle$, because of the quasienergy difference $|\varepsilon_\pi - \varepsilon_0| = \pi/T$. This (quasi)energy splitting is firstly found in $\pi$-spin glass and Floquet time crystals based on MBL [9, 12]. Since the Floquet phase coexistence is protected by the 0 gap and $\pi$ gap separately, the 2T periodic oscillation behaves like an isolated two-level "Rabi" oscillation (Fig. 1g) with eliminating the gapped bulk states and suppressing the scatterings from perturbations [46]. As a consequence, the period-doubling inherits the rigidity from the topological phase coexistence of the Floquet system and breaks the discrete time-translation symmetry at particular frequencies.

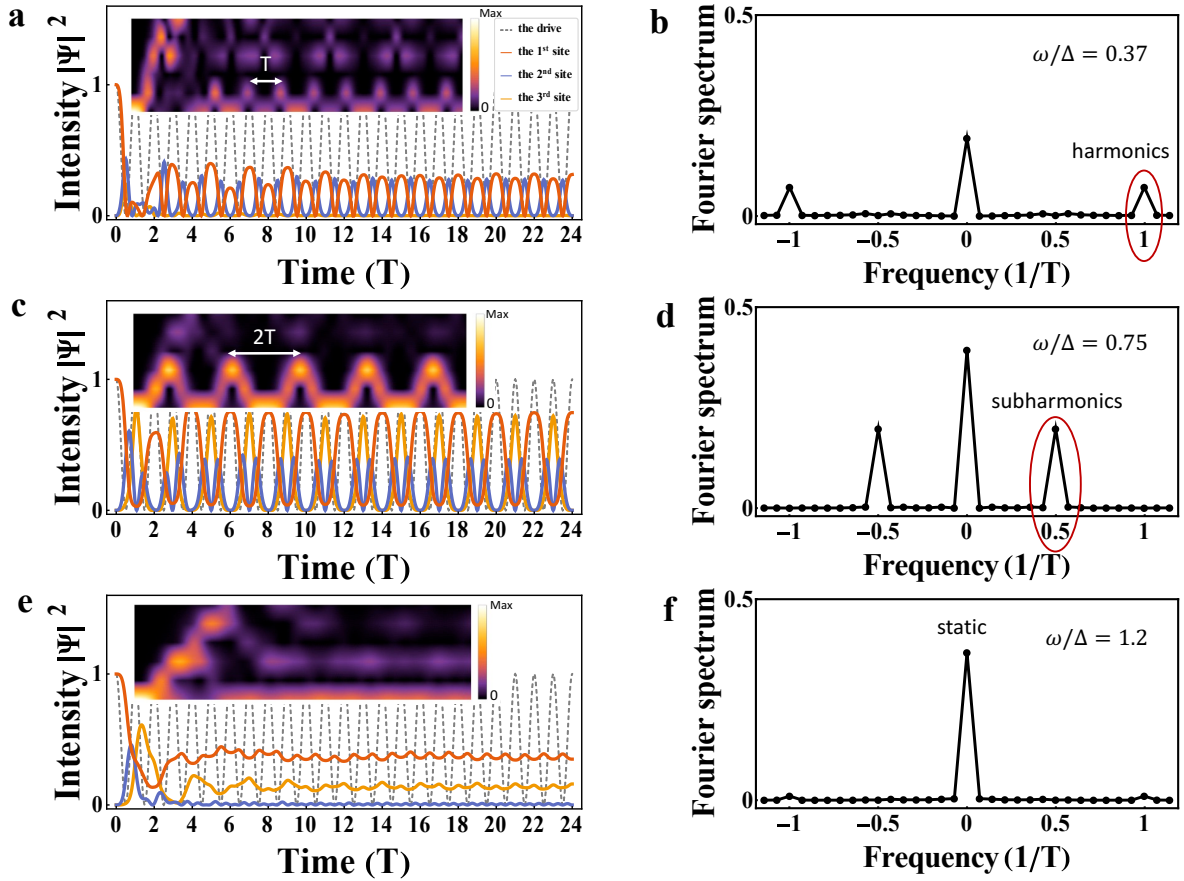

Fig. 3: Demonstrations of period-doubling (period-2T oscillation) in topological phase transition from only $\pi$ modes, topological phase coexistence, to only 0 modes. (a), (c), (e), show the dynamics of the first three sites in different topological driven regions from only $\pi$ modes, phase coexistence, to only 0 modes, respectively, with the only initial input from the first site on the boundary. The insets show the density plot of the stroboscopic dynamical patterns of the zoomed lattice. (b), (d), (f), show the corresponding Fourier spectrum of those dynamics in different regions. (a), (b), the only $\pi$-mode phase at $\omega/\Delta = 0.37$ displays the behavior of the harmonics as following the frequency (1/T) of the drive; (c), (d), the phase coexistence at $\omega/\Delta = 0.75$ displays the subharmonics with period-2T oscillation; (e), (f), the only 0 mode phase at $\omega/\Delta = 1.2$ displays the effectively static behavior with suppression of the harmonics and subharmonics.

We numerically demonstrate the topologically protected period doubling in the topological transition process from only $\pi$-modes phase, two modes coexistence phase, and only 0-modes phase. Clearly, Fig. 3 shows the typical stroboscopic evolutions at different frequencies corresponding to the three topological phases. We numerically calculate the time-dependent wavefunction propagation $\psi(t) = U(t)\psi(0)$, where $\psi(0)$ is the given initial field distribution along the boundary of the crystal. Figs. 3a, 2c, 3e show the calculated intensities $|\psi_i(t)|^2$ on the first three sites ($i = 1, 2, 3$) at certain frequencies $\omega/\Delta = 0.37, 0.75, 1.2$, respectively. The insets demonstrate the density plots of the full intensity $|\psi(t)|^2$ on the partially zoomed lattices (we plotted the first six sites or waveguides from $N = 80$). Only in the region of phase coexistence (Fig. 3c, at $\omega/\Delta = 0.75$), the field intensity unexpectedly oscillates along the boundary ($i = 1$) with two periods (2T), while the rest two patterns of the evolutions (Fig. 3a, 3e) exhibit the conventional stroboscopic behavior undergoing the period of the drive. In the phase with only $\pi$ modes at $\omega/\Delta = 0.37$, the field intensity only propagates between the first and second sites over 24 periods (almost no energy occupied at the third site), whose near-field evolution pattern has been recently observed by Q. Cheng et al. [43]. On the other hand, in the high-frequency approximation region, the 0 mode pattern shows its dominant effective static behavior, as shown in Fig. 3e. The second site ($i = 2$) has no energy occupied due to the emergent sublattice symmetry in the high-frequency limit, which requires the end states only localizes at the odd sites $i = 1, 3$ (or even sites if counting from the other boundary).

To explore the oscillation of the system in a long-range response, we perform the Fourier transformation by taking the stable field intensity $|\psi_1(t)|^2$ on the boundary in time ranging from 10T to 24T. The corresponding Fourier spectrum at different frequencies are presented in Fig. 3b, 3d, 3f, and clearly, we found that the harmonics, subharmonics, and static response dynamically dominates in three different driven Floquet regions, respectively, where only the subharmonic response in the phase coexistence region is rigorously locked at half the drive frequency (1/2T). As a result, we can attribute the period-doubling phenomenon to the topological phase coexistence in the quasienergy spectrum.

**Floquet time crystals in topological phase transition** - The spontaneous breaking of discrete time-translation symmetry indicates that the non-interacting Floquet system is driven into the time-crystalline phase. Fig. 4a exhibits the Fourier components of harmonic, subharmonic, and static response on the boundary in the full frequency range. The only subharmonic component (the red curve) emerges in the frequency region $1/2 < \omega/\Delta < 1$, with the suppression of the harmonic oscillation (the blue curve). Conversely, as shown in Fig. 4a, the harmonic and static components dominate the stroboscopic dynamics of the boundary at the regions related to the appearance of only $\pi$ modes and only 0 modes, respectively. As the drive frequency increases, the harmonic component suddenly comes down to zero from $\omega/\Delta > 1/2$, and the subharmonics plays a leading role in the dynamics of our system where the period-$2T$ oscillation appears. This drive frequency region is called the topological Floquet time-crystalline phase, in accompanying with the broken of discrete time-translation symmetry spontaneously into two-cycles (2T). Finally, at the high-frequency region $\omega/\Delta > 1$, only the static component exists, implying the appearance of only 0 modes as the driven system (Fig. 1d) is mapped into a static SSH system with no drive (Fig. 1b).

To present the Floquet phase transitions more intuitively and directly, we demonstrated the stroboscopic evolution patterns of near-field intensity distribution varying with different drive frequencies, as shown in Fig. 5. In the low-frequency region, only $\pi$ mode can be excited, and the field intensity distribution exhibits harmonic behavior that oscillates with the same period as the drive (Figs. 5a, 5b). As the frequency increases, the system enters the TFTC phase and exhibits stable period-2T oscillation (Figs. 5c-5f). At the high-frequency region, only 0 mode exists, and the driven system manifests the effective high-frequency behavior as same as that of the static SSH model (Fig. 5g and 5h).

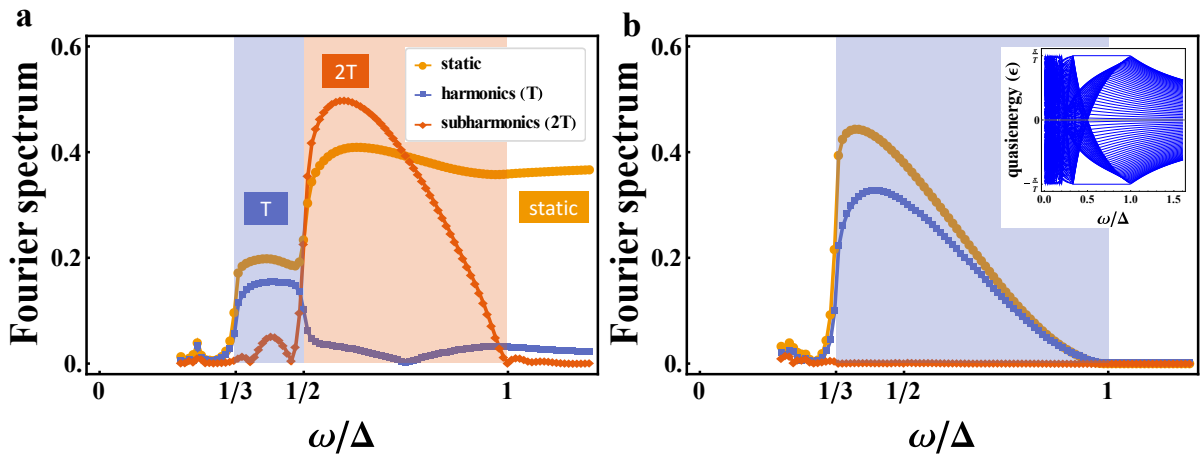

Fig. 4: The static, harmonic, and subharmonic responses of the protected boundary state in topological Floquet time-crystalline (TFTC) phase transition as a function of the drive

frequency $\omega/\Delta$. (a) the harmonic oscillation (T) dominates in the frequency region $(1/3, 1/2)$, the subharmonic oscillation (2T) emerges in the region $(1/2, 1)$ and the static response dominates in the region $(1, \infty)$. The subharmonic response indicates the spontaneous breaking of discrete time-translation symmetry. As a contrast, all these components in the setup of the driven SSH model ($\delta\kappa_0 = 0$) corresponding to Fig.1c is shown in (b). Here, no subharmonic response appears because there is no topological phase coexistence, as shown in the inset of (b), while the harmonics correspond to the sole occurrence of $\pi$ modes.

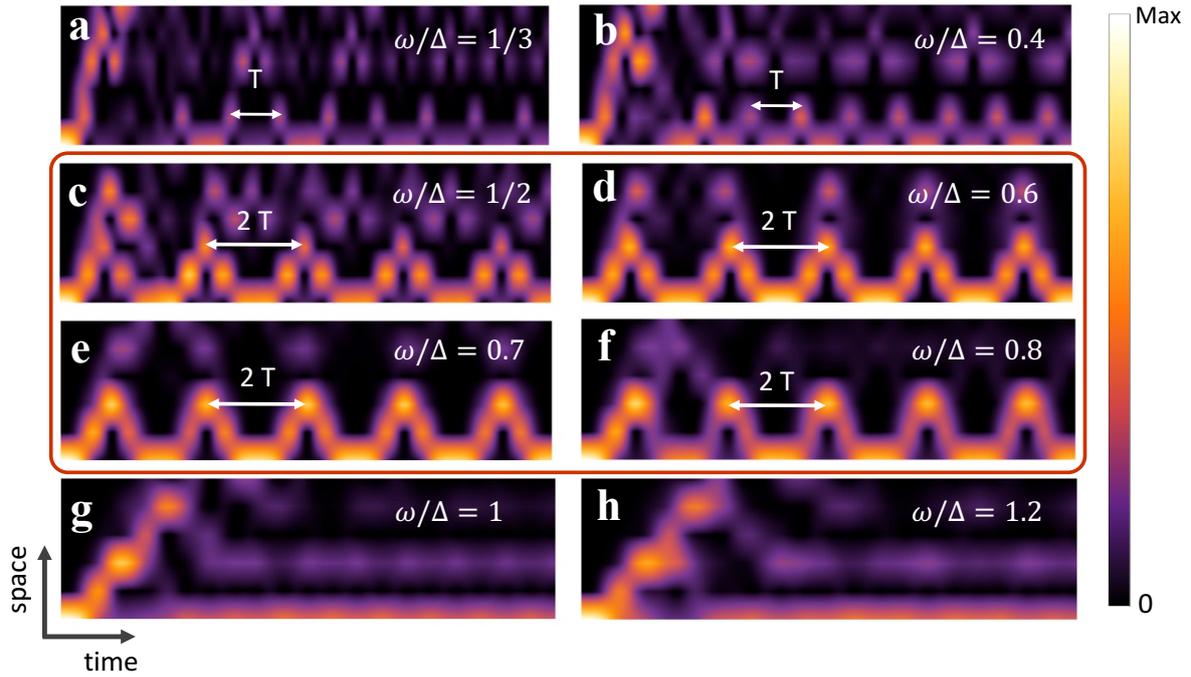

Figure 5: The stroboscopic dynamics of topological Floquet time crystal in different drive frequencies. (a), (b) are in the only $\pi$ mode phase region; (c)- (f) are in the phase coexistence region; (g), (h) are in the only 0 mode phase region. In the topological phase coexistence region (c)-(f), all the dynamics show the period-2T oscillation. The parameters are $N = 80$, $\kappa_0/\Delta = 0.25$, $\delta\kappa_0/\Delta = 0.06$, $\delta\kappa_1/\Delta = 0.12$ and only the first six waveguides are showed.

As a contrast, we also study the characteristic of Fourier components of the stroboscopic dynamics in the driven configuration $\delta\kappa(t) \neq 0, \delta\kappa_0 = 0$ (see the setup in Fig. 1c), in which only $\pi$-mode phase exists in the region $1/3 < \omega/\Delta < 1$. The corresponding quasienergy spectrum is presented in the inset of Fig. 4b, in which the topological $\pi$ modes manifest itself but without coexistence of 0 modes. Fig. 4b shows no subharmonic oscillation in the full driving range. It is worth to address that the harmonic component as a function of frequency has a sharp slope at $\omega/\Delta = 1/3$ but gradually disappears at $\omega/\Delta = 1$. This morphological feature of the harmonic component is correspondingly similar to the lineshape of the $\pi$ gap function in quasienergy spectrum. Inspired by this speculation, we can explain the sharp and slow slopes of the subharmonic component (Fig. 4a) directly rely on the lineshape feature of the emerging 0 gap at $\omega/\Delta = 1/2$ and the disappearing $\pi$ gap at $\omega/\Delta = 1$, respectively.

**Immunization against disorders** - Finally, we assess the robustness against disorders for the subharmonic response of topological Floquet time crystals. Fig. 6 demonstrates the subharmonic components in the phase coexistence region as a function of the strength (W) of on-site disorders (Fig. 6a) and off-diagonal disorders (Fig. 6b) at the frequency $\omega/\Delta = 0.75$. We find that the spectrum peaks survive around at half the drive frequency and become more broad and diffusive when the strength of disorders W is increasingly higher than the emergent 0 and $\pi$ gaps corresponding the staggered coupling strengths $\delta\kappa_0/\Delta = 0.06$, $\delta\kappa_1/\Delta = 0.12$. It means that the stability of our TFTC is comparably rigid against disorders or imperfections in the case that the magnitude of the quasienergy gap is large enough to forbidden the disorder-induced scattering into the bulk states.

Note, that in our driven SSH model, the exact period-doubling in TFTC suffers from local perturbations ($\delta$) on the boundaries because the chiral (sublattice) symmetry is explicitly broken [50]. This is because the chiral symmetry and indivisibility of the unit cell are too fragile by adding an onsite potential. As we will discuss in next section, the Majorana zero modes in Kitaev model behave more robust because the particle-hole symmetry is intrinsic, even though the Kitaev wire can be equivalently mapped to the SSH model. In this situation, admittedly, the subharmonic response is still practically protected by two gaps, but it undergoes a slightly different stable period ($T' \neq 2T$) longer than that of the drive when the perturbation energy $\delta$ is added locally on the boundary (Fig. 6a). Thus, the TFTC phase with a rigid longer-period oscillation ($T' > T$) can safely remain in the presence of chiral (sublattice) symmetry, as verified by our numerical calculations as presented in Fig. 6.

Additionally, the topologically-protected period-doubling can emerge both in the driven edge states and the well-separated driven domain walls. The period-doubling of domain walls emerges in the bulk of the Floquet system, but still robust against perturbations. The distribution of the time-crystalline domain walls can be randomly located in the bulk, which can further suppress the fluctuation and perturbations of disorders and even interactions. We demonstrate the period-2T oscillation in domain walls, as shown in Fig. S2 of the Appendix A3.

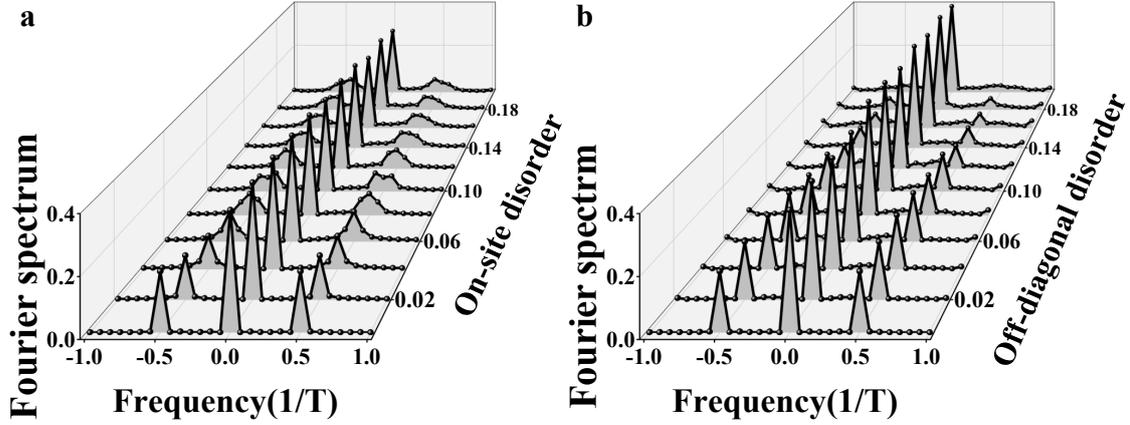

Fig. 6: Robustness against disorders of the emergent subharmonic oscillation in topological Floquet time crystals. The averaged subharmonic component in the Fourier spectrum on the boundary is calculated as a function of the strength of on-site disorder (a) and off-diagonal disorder (b) in the phase coexistence region at $\omega/\Delta = 0.75$. The subharmonic oscillation retains rigidly in small disorders but eventually diminishes as increasing the strength of disorders exceed the emergent energy scale of 0 and $\pi$ gaps in the quasienergy spectrum. The distribution of disorders is random with the unit of the bandwidth $\Delta$. The relevant coupling strengths are $\kappa_0/\Delta = 0.25$, $\delta\kappa_0/\Delta = 0.06$, $\delta\kappa_1/\Delta = 0.12$ and the time window we used to perform the Fourier transformation is $[10T, 24T]$ to avoid counting the components from bulk state scattering.

**Time-crystalline phases in generic one-dimensional driven models**

To demonstrate the ubiquitous existence of topological Floquet time crystal beyond the driven SSH model, here we reconstruct the coexistence of Majorana zero and pi modes in two well-known toy models, that are, Kitaev's toy model for p-wave superconductors [57, 58] and Transverse field Ising model (TFIM) for one-dimensional spin-1/2 chain [9-12], respectively, with involving the Floquet engineering protocols

$$H_{Kitaev}(t) = iJ(t) \sum_{i=1}^{N-1} \gamma_{2i}\gamma_{2i+1} + ih(t) \sum_{i=1}^{N} \gamma_{2i}\gamma_{2i-1},$$
$$H_{TFIM}(t) = -J(t) \sum_{i=1}^{N-1} \sigma_i^z \sigma_{i+1}^z - h(t) \sum_{i=1}^{N} \sigma_i^x, \tag{2}$$

where $\gamma_i$ are Majorana fermions and $\sigma_i^{x,z}$ are Pauli operators. As compared with the driven Su-Schrieffer-Heeger model (Eq.1), we require the parameters $J(t) = \kappa_0 + (\delta\kappa_0 + \delta\kappa(t))$, $h(t) = \kappa_0 - (\delta\kappa_0 + \delta\kappa(t))$ dimerized both spatially and temporally, where $\delta\kappa_0$ is the global dimerization and $\delta\kappa(t) = \delta\kappa(t+T)$ is the time-periodic dimerization. The parameters in Eq.2 have different physical meanings. For instance, $J(t)$ denotes the short-range spin-spin interaction in a spin chain while it denotes the staggered coupling strength in the Kitaev and SSH models. Interestingly, as we already known [57], the Kitaev model connects with our dimerized model through a partial partial-hole mapping, and as well connects with the TFIM model through a non-local Jordan-Wigner transformation. Alternatively, the Floquet topological phases and invariants in these related driven models can be identified by the topological classification in terms of symmetry classes in Floquet systems [59-62].

Many one-dimensional driven models [63-65] (e.g., Eqs. 1 and 2) can be solved likewise, as we investigated in this work, and consequently, they hold typical Floquet topological excitations, such as edge modes and domain walls, which are spanned in sublattice, Nambu (particle-hole) and spin subspaces, respectively. In these driven models with anomalous Floquet topological phases, therefore equivalently, the universal coexistence between 0 and $\pi$ edge modes (or domain walls) can lead to the period-doubling superposition that indicates a class of ubiquitous presence of Floquet time-crystalline period-$2T$ oscillations in one-dimensional periodically driven systems. Regarding the further experimental realization, we address that usually the spin-spin interaction $J(t)$ is assumed to be time-independent for a spin chain and only the transverse field strength $h(t)$ is time-periodic which can be experimentally controlled by exerting an external magnetic field on the spin chain [64]. For realization in p-wave superconductor, the Floquet drive is more flexible to assign on the chemical potential controlled by a time-periodic gate-voltage [44, 45]. It is worthy to note that these three intimately-related TFTCs can be viewed as the hand-waving starting point to engage with various disorders, weak interactions, and other parameters protocols in such as topological insulators and superconductors [35, 66], topological photonics and quantum synthetic materials.

**Conclusion** - To conclude, we reported a period-2T topological Floquet time crystal (TFTC) based on topological phase coexistence between 0 modes and $\pi$ modes in a driven SSH model that exhibits a rigid and long-range subharmonic response, and furthermore, we extended the ubiquitous presence of TFTCs in one-dimensional periodically-driven systems. The rigidity of period-doubling against perturbations is topologically protected by both the nontrivial gap invariants; thus, as a consequence, it requires no many-body interactions or disorders. The mechanism of spontaneous breaking of discrete time-translation symmetry in the Floquet regime of phase coexistence can be experimentally realized in many platforms of topological insulators and topological photonics, such as cold atoms [52, 67, 68] and photonics/microwave platforms [43, 53, 54]. Also, our SSH manifestation of TFTC is a direct connection between topological insulators and Floquet (discrete) time crystals.

To extend, we would like to address two aspects. First, the TFTC phase can also be easily realized in topological superconductors, by experimentally achieving the coexistence between Floquet Majorana $\pi$ modes and Majorana 0 modes [9, 34]. Those experimental platforms have proposed in many literature, such as [44, 45], but unfortunately, without experimentally detecting the 'fingerprint' of the novel period-2T oscillation yet. With the available technical conditions in state-of-the-art experiments, it seems promising to realize a superconducting time crystal in near future. Second, the topological subharmonic response indicates an implementation of subharmonic radiation from the protected boundary/edge states when the interacting electromagnetic field is coupled and detected. The topological-protected subharmonic radiation is promising an exotic optical parametric down-conversion process when the drive is viewed as the "pump" source for the artificially driven materials, which deserves more attention both from theorists and experimentalists.

**Acknowledgement** - We would like to thank the helpful discussions with Binghai Yan, Erez Berg, Netanel Lindner, Qingqing Cheng. This work is supported in parts by ICORE— Israel Center of Research Excellence program of the ISF, and by the Crown Photonics Center.

Y. P. and B. W. contribute equally to this work.

# Appendix

## A1. Quasienergy spectrum from Floquet-Bloch theorem

The Floquet Hamiltonian $H_F$ can be expressed in the matrix form with the elements

$$\langle\langle \alpha, n | H_F | \beta, m \rangle\rangle = \frac{1}{T}\int_0^T \langle \alpha | e^{-in\omega t}(H(t) - i\partial_t)e^{im\omega t} | \beta \rangle dt = \langle \alpha | H^{(n-m)} | \beta \rangle + n\omega \delta_{n,m}\delta_{\alpha,\beta},$$

(S1)

where $|\alpha\rangle$ denotes an orthonormal basis state of the system Hilbert space $\mathcal{H}$ and $|\alpha, n\rangle\rangle$ denotes an orthonormal basis state of the composed Floquet-Hilbert space $\mathcal{S}$[69].

According to (S1), the Floquet Hamiltonian $H_F$ can be expressed as the matrix form

$$H_F = \begin{pmatrix} \ddots & \vdots & \vdots & \vdots & \vdots & \vdots & \reflectbox{$\ddots$} \\ \cdots & H^{(0)} - 2\omega & H^{(-1)} & 0 & 0 & 0 & \cdots \\ \cdots & H^{(1)} & H^{(0)} - \omega & H^{(-1)} & 0 & 0 & \cdots \\ \cdots & 0 & H^{(1)} & H^{(0)} & H^{(-1)} & 0 & \cdots \\ \cdots & 0 & 0 & H^{(1)} & H^{(0)} + \omega & H^{(-1)} & \cdots \\ \cdots & 0 & 0 & 0 & H^{(1)} & H^{(0)} + 2\omega & \cdots \\ \reflectbox{$\ddots$} & \vdots & \vdots & \vdots & \vdots & \vdots & \ddots \end{pmatrix},$$

(S2)

where the matrix block $H^{(n)}$ is the $n^{th}$ Fourier component of $H(t)$. Here, the diagonal terms, $n = 0, \pm 1, \pm 2 \ldots$ define the artificial photon bands with photon energy $n\omega$ and the decoupled local electron energy given by $H^{(0)}$. The secondary-diagonal terms $H^{(\pm 1)}$ correspond to the artificial photon scattering between two neighbor photon bands $n$ and $n + 1$, where the sign "-" implies the emission process and the sign "+" implies the absorption process. In our Floquet protocol of driven SSH model, all the high-order photon scattering $H^{(\pm l)}$ ($l \geq 2$) are suppressed.

In Wannier representation with considering the open boundary condition (or periodic boundary condition), the Hamiltonian components are given by ($N$ is an even number)

$H^{(0)} =$

$$\begin{pmatrix} \beta_0 & \kappa_0 - \delta\kappa_0 & 0 & 0 & 0 & 0 \text{ (or } \kappa_0 + \delta\kappa_0) \\ \kappa_0 - \delta\kappa_0 & \beta_0 & \kappa_0 + \delta\kappa_0 & 0 & 0 & 0 \\ 0 & \kappa_0 + \delta\kappa_0 & \beta_0 & \ddots & 0 & 0 \\ 0 & 0 & \ddots & \ddots & \kappa_0 + \delta\kappa_0 & 0 \\ 0 & 0 & 0 & \kappa_0 + \delta\kappa_0 & \beta_0 & \kappa_0 - \delta\kappa_0 \\ 0 \text{ (or } \kappa_0 + \delta\kappa_0) & 0 & 0 & 0 & \kappa_0 - \delta\kappa_0 & \beta_0 \end{pmatrix}_{N\times N},$$

(S3)

$$H^{(1)} =$$

$$\begin{pmatrix} 0 & -\frac{\delta\kappa_1}{2}e^{i\theta} & 0 & 0 & 0 & 0(or\frac{\delta\kappa_1 e^{i\theta}}{2}) \\ -\frac{\delta\kappa_1}{2}e^{i\theta} & 0 & \frac{\delta\kappa_1}{2}e^{i\theta} & 0 & 0 & 0 \\ 0 & \frac{\delta\kappa_1}{2}e^{i\theta} & 0 & \ddots & 0 & 0 \\ 0 & 0 & \ddots & \ddots & \frac{\delta\kappa_1}{2}e^{i\theta} & 0 \\ 0 & 0 & 0 & \frac{\delta\kappa_1}{2}e^{i\theta} & 0 & -\frac{\delta\kappa_1}{2}e^{i\theta} \\ 0\ (or\frac{\delta\kappa_1 e^{i\theta}}{2}) & 0 & 0 & 0 & -\frac{\delta\kappa_1}{2}e^{i\theta} & 0 \end{pmatrix}_{N\times N},$$

(S4)

$$H^{(-1)} =$$

$$\begin{pmatrix} 0 & -\frac{\delta\kappa_1}{2}e^{-i\theta} & 0 & 0 & 0 & 0\ (or\frac{\delta\kappa_1 e^{-i\theta}}{2}) \\ -\frac{\delta\kappa_1}{2}e^{-i\theta} & 0 & \frac{\delta\kappa_1}{2}e^{-i\theta} & 0 & 0 & 0 \\ 0 & \frac{\delta\kappa_1}{2}e^{-i\theta} & 0 & \ddots & 0 & 0 \\ 0 & 0 & \ddots & \ddots & \frac{\delta\kappa_1}{2}e^{-i\theta} & 0 \\ 0 & 0 & 0 & \frac{\delta\kappa_1}{2}e^{-i\theta} & 0 & -\frac{\delta\kappa_1}{2}e^{-i\theta} \\ 0\ (or\frac{\delta\kappa_1 e^{-i\theta}}{2}) & 0 & 0 & 0 & -\frac{\delta\kappa_1}{2}e^{-i\theta} & 0 \end{pmatrix}_{N\times N},$$

(S5)

and $H^{(\pm n)} = 0$ for $n \geq 2$. Then, we can get the quasienergy spectrum as exhibited in Fig. 2(truncated for five Floquet replicas) from the secular equation

$$\left|H_F^{(trucated)} - \varepsilon I\right| = 0.$$  (S6)

Because the spanned Floquet-Hilbert space is infinite, as shown in (S2), the practical calculation requires truncation in order to force the Hilbert into an effective finite-dimension space. The general truncation, in principle, is complicated. In our driven SSH case, however, only the first-order photon scattering is relevant, thus we truncation the space into five photon bands as calculated numerically in Fig. 2.

To explicitly explore the gap opening mechanism of the isolated modes, here let us represent our driven SSH Hamiltonian in the Bloch basis. Considering the periodic boundary condition, due to the translation symmetry, we can transform the Eq.1 in the main text into the momentum space with the corresponding Bloch representation as [57]

$$H(k,t) = \left(\kappa_0 - \delta\kappa_0 - \delta\kappa(t) + \left(\kappa_0 + \delta\kappa_0 + \delta\kappa(t)\right)\cos(k)\right)\sigma_x + \left(\kappa_0 + \delta\kappa_0 + \delta\kappa(t)\right)\sin(k)\,\sigma_y, \tag{S7}$$

where $\sigma_x, \sigma_y$ are the Pauli matrices on the basis of sublattices A and B. These Pauli operators are used to characterize the sublattice space corresponding to sublattice symmetry (i.e., chiral symmetry) after dimerization. Notice that the anti-commutation relation $\{H(k,t), \sigma_z\} = 0$ indicates that our driven SSH modeling of period-2T Floquet time crystal possesses a sublattice ('chiral') symmetry defined by the Pauli operator $\sigma_z$ [43]. In this case, the matrix blocks of the Floquet Hamiltonian $H_F$ in the analytical 2×2 matrix form can be written as

$$H^{(0)} = [(\kappa_0 - \delta\kappa_0) + (\kappa_0 + \delta\kappa_0)\cos(k)]\sigma_x + [(\kappa_0 + \delta\kappa_0)\sin(k)]\sigma_y, \tag{S8}$$

$$H^{(1)} = \left[-\frac{\delta\kappa_1 e^{i\theta}}{2} + \frac{\delta\kappa_1 e^{i\theta}}{2}\cos(k)\right]\sigma_x + \frac{\delta\kappa_1 e^{i\theta}}{2}\sin(k)\,\sigma_y, \tag{S9}$$

$$H^{(-1)} = \left[-\frac{\delta\kappa_1 e^{-i\theta}}{2} + \frac{\delta\kappa_1 e^{-i\theta}}{2}\cos(k)\right]\sigma_x + \frac{\delta\kappa_1 e^{-i\theta}}{2}\sin(k)\,\sigma_y. \tag{S10}$$

Still, $H^{(\pm n)}(k) = 0$ for $n \geq 2$. The quasienergy spectrum (truncated for five Floquet replicas) concerning quasi-momentum $k$ is presented in Fig. S1 which reveals the gap "open-close-open" mechanism.

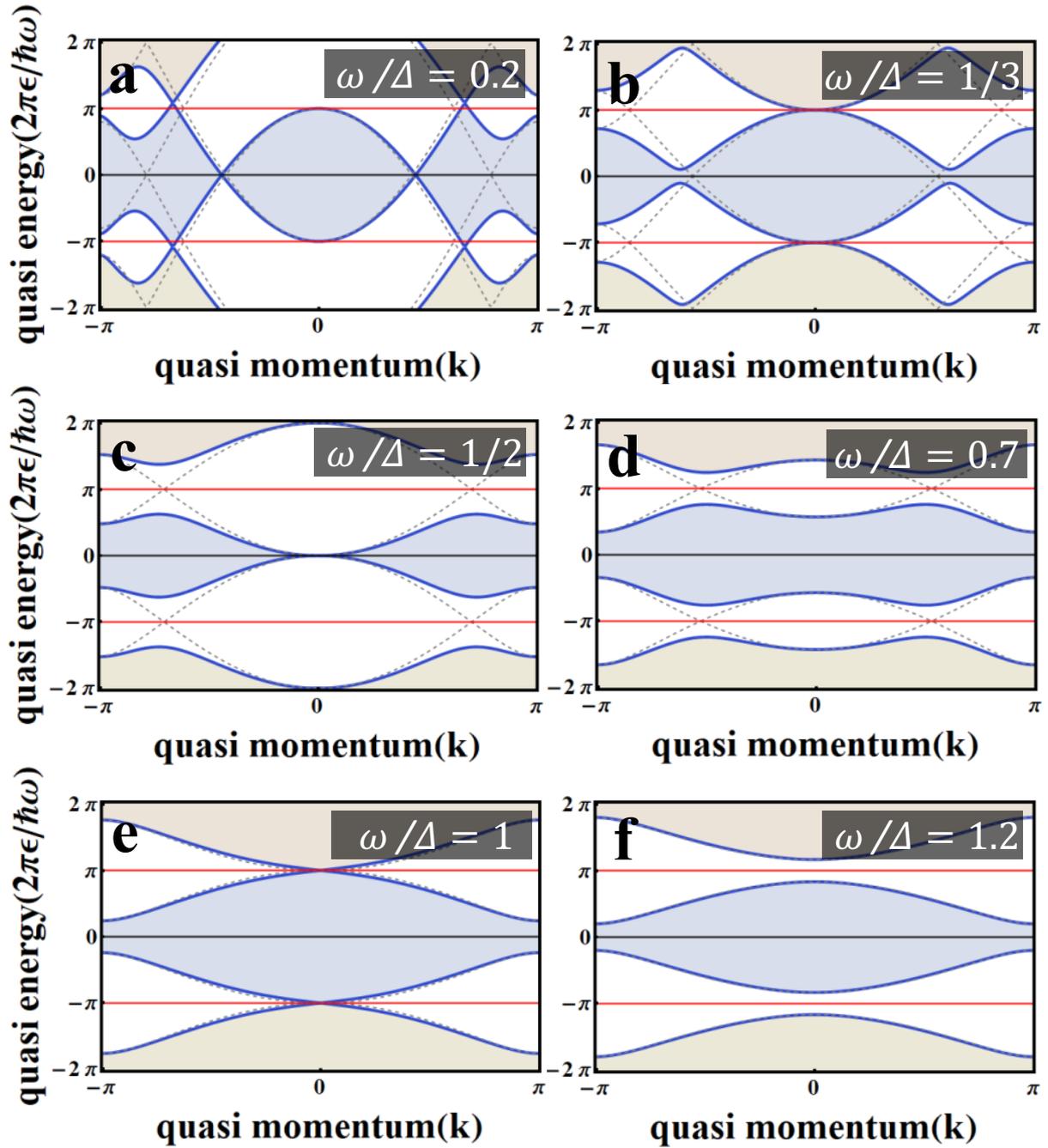

Figure S1: The "open-close-open" mechanism of 0 gap and $\pi$ gap in topological Floquet time crystals. Only in the region $(1/2, 1)$ both the 0 and $\pi$ gaps open nontrivially.

At high frequency, different photon bands are entirely separated, and the 0 gap and $\pi$ are both open. As the frequency decreases, photon bands come to overlap with each other. At $\omega/\Delta = 1$, the $\pi$ gap closes due to the mixing of the $n = 0$ and $n = \pm 1$ photon bands. At $\omega/\Delta = 1/2$, the $n = 1$ and $n = -1$ photon bands cross, and the 0 gap closes. When decreasing the driving frequency to $\omega/\Delta = 1/3$ where $n = 2$ and $n = -1$ ($n = -2$ and $n = 1$) photon bands crossing, the $\pi$ gap closes. The $\pi$ mode gap starts to open at $\omega/\Delta = 1/3$ and close at $\omega/\Delta = 1$ while the 0 gap "open-close-open" transition occurs at $\omega/\Delta = 1/2$. The 0 and $\pi$ gaps keep open simultaneously in the frequency region $1/2 < \omega/\Delta < 1$, which is consistent with the above result.

**A2. The effective Hamiltonian from time-periodic evolution operator**

Now, we calculate the evolution of the periodically-driven system with a given initial condition. The dynamic process of a time-periodic system of which the Hamiltonian satisfies $H(T + t) = H(t)$ can be described by the time evolution operator

$$|\psi(t)\rangle = U(t, t_0)|\psi(t_0)\rangle, \tag{S11}$$

where $|\psi(t)\rangle$ denotes the state of the system at time $t$, and $|\psi(t_0)\rangle$ denotes the initial state at time $t_0$. The time evolution operator $U(t, t_0)$ is determined by the differential equation

$$i\partial_t U(t, t_0) = H(t)U(t, t_0), \tag{S12}$$

which has the solution

$$U(t, t_0) = \hat{T} e^{-i \int_{t_0}^{t} H(t')dt'}, \tag{S13}$$

with an initial value $U(t, t_0) = Id$ where $\hat{T}$ is the time-ordering operator. Note that the discrete time-translation symmetry $U(t_2, t_1) = U(t_2 + T, t_1 + T)$ and $U(t_2, t_1) = U(t_2, t')U(t', t_1)$ for arbitrary time $t'$ in between the interval $(t_1, t_2)$, the dynamic evolution operator from time $t_0$ to $t$, can be rewritten as

$$U(t, t_0) = U(t, t_0 + nT)U(t_0 + nT, t_0)$$
$$= U(t, t_0 + nT)[U(t_0 + T, t_0)]^n, \tag{S14}$$

where n is an integer and the time interval $\delta t = (t - t_0 - nT) \in [0, T]$.

From (S14), we can see that the stroboscopic evolution observed at each period can be fully described by $U(t_0 + T, t_0)$ called the Floquet operator [40] which satisfies

$$U(t_0 + T, t_0)|\psi(t_0)\rangle = e^{-i\varepsilon T}|\psi(t_0)\rangle, \tag{S15}$$

where $|\psi(t_0)\rangle$ is the Floquet state, and $\varepsilon$ is the quasienergy. It is convenient to define the evolution within one period as an evolution with the time-independent effective Hamiltonian

$$U(t_0 + T, t_0) = e^{-iH^{eff}[t_0]T}, \tag{S16}$$

where the choice of $t_0$ is arbitrary, which is called the Floquet gauge [39]. For example, in the main text, the driving field is $\cos\omega t$; we choose $t_0 = 0$, which is the symmetric point of the driving protocol. In this case, the effective Hamiltonian is defined by

$$H_\eta^{eff} = \frac{i}{T}\log_{-\eta} U(T), \tag{S17}$$

where for simplification we suppress the gauge index $t_0$ such that $H_\eta^{eff} = H_\eta^{eff}[t_0 = 0]$, $U(T) = U(T,0)$ and we notice that the function $\log_\eta$ is the complex logarithm with branch cut along an axis with angle η, defined as [42]

$$\log_{-\eta} e^{i\varphi} = i\varphi, \quad \text{for } -\eta - 2\pi < \varphi < -\eta. \tag{S18}$$

The Floquet operator $U(T)$ and the effective $H_\eta^{eff}$ share the same eigenstates and

$$H_\eta^{eff}|\psi\rangle = \varepsilon_\eta|\psi\rangle, \tag{S19}$$

where $\varepsilon_\eta = \frac{i}{T}\log_{-\eta}\varepsilon$ is the quasienergy.

Then we can get the quasienergy spectrum (here $\eta = \pi$) of the effective Hamiltonian exhibited in Fig. 1e, and Fig. 3b in the main text. The dynamic evolution of the time-dependent system can be calculated numerically by the discretized evolution operator

$$U(t) = \lim_{\Delta t \to 0} e^{-iH(t-\Delta t)\Delta t} e^{-iH(t-2\Delta t)\Delta t} \dots e^{-iH(\Delta t)\Delta t} e^{-iH(0)\Delta t}, \tag{S20}$$

Up to now, we have given two equivalent methods to calculate the quasienergy spectrum by the Floquet Hamiltonian $H_F$ in the differential form and the effective Hamiltonian $H_\eta^{eff}$ in the integral form, respectively. Even though they are comparing different in its forms, we should be able to find the unitary transformation to prove the isomorphic equivalence between their spectrum. For this reason, let us return the stroboscopic evolution operator as given in (S16) and substitute it into (S14), we obtain

$$\begin{aligned} U(t, t_0) &= U(t, t_0 + nT)[U(t_0 + T, t_0)]^n \\ &= U(t, t_0 + nT)e^{iH_\eta^{eff}[t_0](t-t_0-nT)} e^{-iH_\eta^{eff}[t_0](t-t_0-nT)} e^{-iH_\eta^{eff}[t_0]nT} \\ &= V_\eta(t, t_0)e^{-iH_\eta^{eff}[t_0](t-t_0)}, \end{aligned} \tag{S21}$$

where the periodized evolution operator $V_\eta(t, t_0) = V_\eta(t + nT, t_0) = V_\eta(t, t_0 + nT)$ is defined as $V_\eta(t, t_0) = U(t, t_0)e^{iH_\eta^{eff}[t_0](t-t_0)}$ that contains the short-time scale information (i.e. the micromotion), while the $H_\eta^{eff}[t_0]$ contains the long-time scale information (i.e., stroboscopic evolution). The Floquet Hamiltonian $H_F$ can be transformed to time-independent effective Hamiltonian $H_\eta^{eff}$ by the unitary rotation $V_\eta(t)$

$$V_\eta^{-1}(t)H_F(t)V_\eta(t) = V_\eta^{-1}(t)(H(t) - i\partial_t)V_\eta(t) = H_\eta^{eff}. \tag{S22}$$

## A3. Period-2T oscillation in driven domain walls

The Peiod-2T oscillation we realized in the main text can not only exist at the boundary of the topological Floquet system, but also in the bulk of the system as the existence of domain walls. As compared, we construct the domain wall structure based on the driven SSH model and calculate the time evolution of the system with different initial state input. Fig. S2 shows the Period-2T oscillation of driven domain walls in the bulk of the Floquet system.

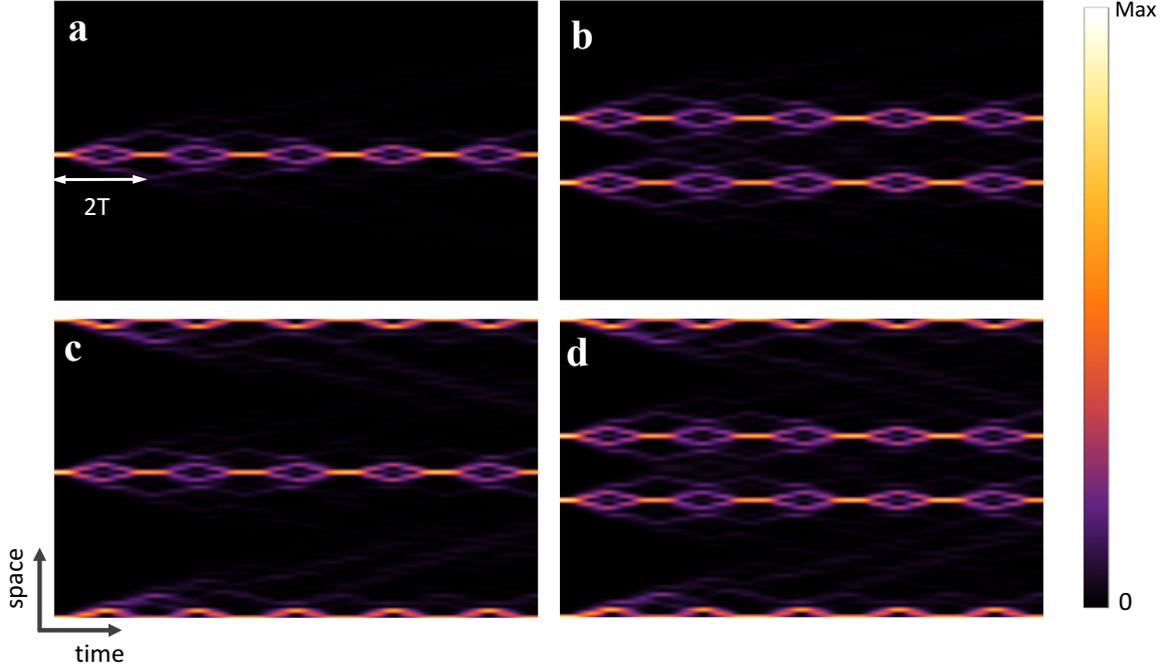

Fig. S2: Period-2T oscillations in driven domain walls. The Floquet structures are based on the driven SSH model with kink configurations in the bulk. (a) Period-2T oscillation of single domain wall with the initial input from the middle domain wall; (b) Period-2T oscillation of two domain walls with the initial input from two well-separated domain walls; (c) Period-2T oscillation of both the domain wall and the edge states on the boundaries of the Floquet system. We obtain the period-doubling simultaneously at the domain wall and two edges; (d) Period-2T oscillation in two well-separated domain walls and the boundaries of the Floquet system. We obtain the period-doubling simultaneously at the two domain walls and two edges. (a), (c) are calculated with 81 waveguides and (b), (d) are 80 waveguides coupled in an array. The relevant parameters are $\kappa_0/\Delta = 0.25, \delta\kappa_0/\Delta = 0.06, \delta\kappa_1/\Delta = 0.12, \omega/\Delta = 0.75$ and the total evolution time is 10T for all the situations.

## A4. The chiral gap invariants in driven SSH model

Consider the static SSH model

$$H(k) = \boldsymbol{h} \cdot \boldsymbol{\sigma} = (\kappa_0 + \delta\kappa_0 + (\kappa_0 - \delta\kappa_0)\cos(k))\sigma_x + (\kappa_0 - \delta\kappa_0)\sin(k)\,\sigma_y, \quad (S23)$$

where $\sigma_x$, $\sigma_y$ are the Pauli matrices in the basis of sublattices A and B. We notice that this Hamiltonian has chiral symmetry which is defined as the unitary chiral operator $\Gamma = \sigma_z$

$$\Gamma H(k) \Gamma^{-1} = -H(k), \quad (S24)$$

When the Hamiltonian is gapped ($\boldsymbol{h} \neq 0$), we can define the valence projector associated with the gapped ground state [57]

$$P = \frac{1}{2}(1 - \boldsymbol{n} \cdot \boldsymbol{\sigma}), \quad (S25)$$

where $\boldsymbol{n} = \boldsymbol{h}/||\boldsymbol{h}||$ and the corresponding unitary operator

$$Q = Id - 2P = \boldsymbol{n} \cdot \boldsymbol{\sigma}, \quad (S26)$$

which is anti-diagonal in the chiral basis

$$Q = \begin{pmatrix} 0 & q \\ q^\dagger & 0 \end{pmatrix}. \quad (S27)$$

The chiral invariant is defined as

$$g = \frac{i}{2\pi} \int_{BZ} dk\, q^{-1}(k) \partial_k q(k), \quad (S28)$$

where $q(k) = n_x - i n_y$.

and for the time-independent SSH model (assume $\kappa_0 > 0$)

$$g = \begin{cases} 1, & \text{when } \delta\kappa_0 < 0, \\ 0, & \text{when } \delta\kappa_0 > 0. \end{cases} \quad (S29)$$

When $g = 1$, the system is topologically nontrivial, which has a topological edge mode while it is trivial for $g = 0$.

However, for time-dependent systems, the invariant developed in the static system do not uniquely determine the number of chiral edge modes within each bulk band gap [49]. Consider the periodically driven SSH model (Fig. 1d),

$$H(k,t) = \left(\kappa_0 + \delta\kappa_0 + \delta\kappa(t) + (\kappa_0 - \delta\kappa_0 - \delta\kappa(t))\cos(k)\right)\sigma_x + (\kappa_0 - \delta\kappa_0 - \delta\kappa(t))\sin(k)\,\sigma_y, \quad (S30)$$

and the Hamiltonian satisfies

$$H(k,t) = H(k, t+T), \quad (S31)$$

where $\delta\kappa(t) = \delta\kappa_1 \cos(\omega t)$ and $T = 2\pi/\omega$ is the driving period.

There is also the chiral symmetry

$$\Gamma H(t,k)\Gamma^{-1} = -H(-t,k). \quad (S32)$$

A $\mathbb{Z}$-valued bulk gap invariant $G_\epsilon[U] \in \mathbb{Z}$ defined for the chiral gaps $\epsilon = 0$ or $\pi$ can work in the driven system with chiral symmetry [42]. Note that the chiral symmetry has the constraint on the periodized evolution operator

$$\Gamma V_\varepsilon(t,k)\Gamma^{-1} = -V_{-\varepsilon}(-t,k)e^{2\pi i t/T}. \quad (S33)$$

Therefore, there are two gap invariants. First, for $\epsilon = 0$ and $t = T/2$, it turns out

$$\Gamma V_0(T/2, k)\Gamma^{-1} = -V_0(T/2, k) \quad (S34)$$

which is anti-diagonal in the chiral basis,

$$V_0(T/2, k) = \begin{pmatrix} 0 & V_0^+ \\ V_0^- & 0 \end{pmatrix}. \quad (S35)$$

The chiral invariant for 0 gap is defined by

$$G_0 = \frac{i}{2\pi} \int_{-\pi}^{\pi} tr((V_0^+)^{-1} \partial_k V_0^+) dk. \quad (S36)$$

Second, for $\epsilon = \pi$ and $t = T/2$, (S33) turns out

$$\Gamma V_\pi(T/2, k)\Gamma^{-1} = V_\pi(T/2, k), \quad (S37)$$

which is diagonal in the chiral basis.

$$V_\pi(T/2, k) = \begin{pmatrix} V_\pi^+ & 0 \\ 0 & V_\pi^- \end{pmatrix}. \quad (S38)$$

The chiral invariant for $\pi$ gap is defined by

$$G_\pi = \frac{i}{2\pi} \int_{-\pi}^{\pi} tr((V_\pi^+)^{-1} \partial_k V_\pi^+) dk. \quad (S39)$$

We can see the chiral invariants as presented in Fig. 1f of the main text, convincing that the 0 mode exists in the high frequency region $\omega/\Delta > 1/2$ correspondingly with the nontrivial 0 gap invariant $G_0 = 1$ while the $\pi$ mode exists in the intermediate frequency region $1/3 < \omega/\Delta < 1$ with the nontrivial $\pi$ gap invariant $G_\pi = 1$, respectively.